\documentstyle[epsf]{lamuphys}
\makeatletter
\let\chapter\hid@chapter
\makeatother
\newcounter{bibcnt}
\newcommand{\bibi}[1]{\bibitem{[\thebibcnt]}{#1}{[\thebibcnt]}\stepcounter{bibcnt}}
\begin{document}
\pagenumbering{arabic}
\title{Transport and Noise of Entangled Electrons}

\author{Eugene V.\,Sukhorukov, Daniel Loss, and Guido Burkard }

\institute{Department of Physics and Astronomy, University of Basel,\\
Klingelbergstrasse 82, CH-4056 Basel, Switzerland}

\maketitle

\renewcommand{\thefootnote}{\fnsymbol{footnote}}
\footnotetext[4]{To appear in the Proceedings of the XVI Sitges Conference, 
Statistical and Dynamical Aspects of Mesoscopic Systems, 
Lecture Notes in Physics, Springer.}
\renewcommand{\thefootnote}{\arabic{footnote}}

\begin{abstract}

We
consider a scattering set-up with an entangler and beam splitter where the
current
noise exhibits bunching behavior for electronic singlet states and
antibunching
behavior for triplet states.
We show that the entanglement of two electrons
in the double-dot  
can be detected in mesoscopic
transport measurements. In the cotunneling
regime
the singlet and
triplet states lead to phase-coherent current contributions of
opposite
signs and to Aharonov-Bohm and
Berry phase oscillations in response
to magnetic
fields.
We analyze the Fermi liquid effects in the transport of entangled electrons.
\end{abstract}

\section{Introduction}
The availability of pairwise entangled qubits -- Einstein-Podolsky-Rosen
(EPR)
pairs \cite{Einstein} -- is a necessary prerequisite in quantum
communication \cite{Bennett84}. The prime example of an EPR pair
considered here  is
the singlet/triplet state formed by two electron spins \cite{Loss98,Burkard}.
Its main feature is
its non-locality: If we separate the two
electrons  from each other in real space, their total spin state
can still remain entangled. Such non-locality gives rise to
striking
phenomena such as violations of Bell inequalities and quantum
teleportation
and has been investigated for photons \cite{Aspect,Zeilinger},
but not yet for {\it massive} particles such as
electrons, let alone in a solid state environment. 
In this work we discuss specific properties of 
transport and noise of entangled electrons as a result of 
two-particle coherence and nonlocality.
 
\begin{figure}[thbp]
\begin{center}
    \leavevmode
\epsfxsize=10.0cm
\epsffile{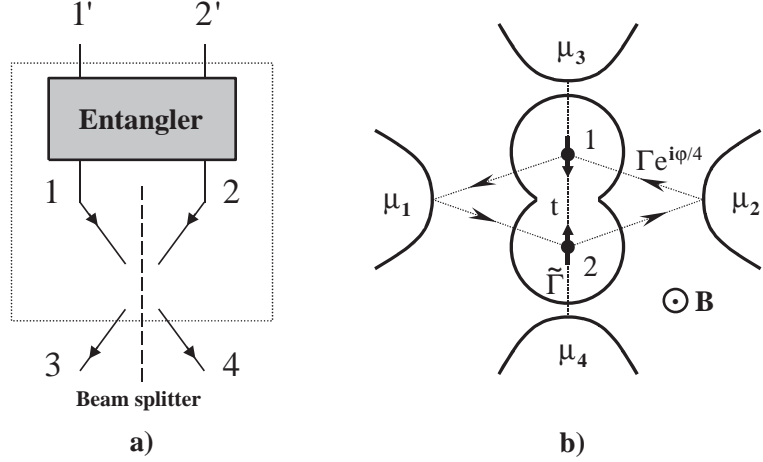}
  \end{center}
\caption{
a) The setup for measuring noise correlations of entangled states.
Uncorrelated electrons are fed into the entangler through the Fermi
leads $1'$ and $2'$. The entangler
is a device  that produces pairs of entangled electrons 
and injects one of the electrons into lead $1$ and the other into
lead $2$. 
The entanglement can then be detected
by performing an interference experiment using a beam splitter. 
b) Double-dot (DD) system containing two electrons and
being weakly coupled to metallic leads 1,...,4, each of which being at the
chemical
potential
$\mu_1$,...,$\mu_4$. The tunneling amplitudes between dots and leads are
denoted
by $\Gamma$, ${\tilde \Gamma}$. The tunneling (t) between the dots
results in a singlet (triplet) ground state.
The closed tunneling path
between dots
and leads 1 and 2 encloses the area $A$.}
\label{fig}
\end{figure}

In Sect. (2) we propose and analyze an experimental set-up 
(see Fig. \ref{fig}a) by which the
entanglement of mobile electrons  can be detected in noise
measurements with a beam splitter \cite{BLS}.  
The entangler is assumed to be a device by which we
can generate entangled electron states, a specific realization
being the double-dot system \cite{Loss98}.  The presence of a beam
splitter ensures that the electrons leaving the entangler have a finite
amplitude to be interchanged.
Thus we can expect that 
the current-current correlations (noise) measured in  leads 3 and/or 4
are  sensitive to the symmetry of the 
{\it orbital part} of the wave function  \cite{Feynman}.
\noindent
Since  the spin
singlet of two electrons is uniquely associated with a symmetric orbital
wave-function, and the triplet with an antisymmetric
one we have a means to distinguish singlets from triplets through a
bunching or antibunching signature.
It is well-known \cite{Loudon} 
that bosons (fermions) show bunching (antibunching) behavior \cite{noise}.
Antibunching is so
far considered for electrons in the normal state both in
theory \cite{Buettiker1,Martin} and in experiments \cite{Stanford}.
However, this classical effect is independent of phase coherence \cite{SL}
and should be carefully distinguished from the two-particle phase-coherent effect 
which we propose here.

The  scheme we propose in Sect. (3)  \cite{LS} consists
of two coupled quantum dots (DD) which
themselves are weakly coupled  in parallel to two leads 1 and 2 (see Fig. \ref{fig}b).  
This results in
a closed loop, and applying a magnetic field, an Aharonov-Bohm (AB) phase
$\varphi$
will be accumulated by an electron traversing the DD. In the Coulomb
blockade
(CB) regime we find that due to
cotunneling \cite{averinazarov} 
the current  depends on the state of the DD:
the AB oscillations for singlet and triplets have opposite sign.
The amplitude of the AB oscillations provides a  measure of
the phase coherence of the entangled state, while the period -- via the
enclosed area
A -- provides a  measure  of the non-locality of the EPR pairs.
The triplets themselves can  be further distinguished by applying a directionally
inhomogeneous magnetic field which adds a Berry phase \cite{LossBerry}
leading to beating.

Finally, in Sect. (4) we would like to address the following 
question \cite{DiVincenzoLossMMM}: Is it
possible to use mobile electrons, prepared in a entangled
spin state, for the purpose of quantum communication? 
Without spin-dependent interaction we know that the total
spin must be conserved even if the two electrons strongly interact
with the other electrons in the mesoscopic environment 
(and among themselves) via Coulomb interaction.  It is
thus not unreasonable to expect that we still find some spin
correlations between initial and final states. But how much is it? 
And why and how do we loose some of the correlations? 

\section{ Noise of Entangled Electrons: Beam Splitter Set-up}

Below, we extend the standard scattering matrix approach \cite{Buettiker1} 
to a situation with entanglement. 
We start by writing the operator for the current carried by electrons
with spin $\sigma$ in lead $\alpha$ of a multiterminal conductor as
\begin{eqnarray}
  I_{\alpha\sigma}(t) &=& \frac{e}{h\nu}\sum_{\epsilon,\epsilon '} \sum_{\beta\gamma}
a_{\beta\sigma}^\dagger(\epsilon) a_{\gamma\sigma}(\epsilon ')
A_{\beta\gamma}^\alpha \exp\left[\I (\epsilon -\epsilon ')t/\hbar\right] \enspace ,\label{current}\\
A_{\beta\gamma}^{\alpha} &=&
\delta_{\alpha\beta}\delta_{\alpha\gamma}
-s_{\alpha\beta}^{*} s_{\alpha\gamma} \enspace .\label{A2}
\end{eqnarray}
where $a^\dagger_{\alpha \sigma} (\epsilon)$ creates an incoming electron in lead
$\alpha$ with spin $\sigma$ and energy $\epsilon$, and  we
assume that the scattering matrix $s_{\alpha\beta}$ is spin- and 
energy-independent.
Note that since we are dealing with discrete energy states here, we
normalize
the operators $a_\alpha(\epsilon)$
such that
$[a_{\alpha\sigma}(\epsilon),a_{\beta\sigma'}(\epsilon ')^\dagger]=
\delta_{\sigma\sigma'}\delta_{\alpha\beta}\delta_{\epsilon,\epsilon '}/\nu$,
where the Kronecker symbol $\delta_{\epsilon,\epsilon '}$ equals $1$ if 
$\epsilon=\epsilon '$ and $0$ otherwise.
Therefore we also have included the factor $1/\nu$ in the definition of
the
current, where $\nu$ stands for the density of states in the leads.
We will also assume that each lead consists of only a single quantum
channel;
the generalization to leads with several channels is straightforward but is
not
needed here. 

We restrict ourselves here to unpolarized currents,
$I_\alpha=\sum_\sigma I_{\alpha\sigma}$.
The spectral density current fluctuations (noise)
$\delta I_{\alpha}=I_{\alpha}-\langle I_{\alpha}\rangle$
between the leads $\alpha$ and $\beta$ are defined as
\begin{equation}
  \label{cross1}
  S_{\alpha\beta}({\omega})
  = \lim_{T\rightarrow\infty}
  \frac{h\nu}{T}\int_0^T\!\!\! d t\,\exp\left(\I \omega t\right)
  \langle\Psi|\delta I_{\alpha}(t)\delta I_{\beta}(0)|\Psi\rangle \enspace ,
\end{equation}
where $|\Psi\rangle$ is the quantum state of the system to be specified
next\footnote{
Note that since $|\Psi\rangle$ in general does not describe a Fermi liquid
state, it is not possible to apply Wick's theorem.}.
We will now investigate the noise  for scattering with the
entangled incident state
\begin{equation}
  \label{entangled_state}
  |\pm\rangle
  = \frac{1}{\sqrt{2}}\left( a_{2\downarrow}^\dagger(\epsilon_2)
a_{1\uparrow}^\dagger(\epsilon_1)
    \pm a_{2\uparrow}^\dagger(\epsilon_2)
a_{1\downarrow}^\dagger(\epsilon_1)\right) |0\rangle \enspace .
\end{equation}
The state $|-\rangle$ is the spin singlet, $|S\rangle$,
while $|+\rangle$ denotes the $S_z=0$  triplet
$|T_{0}\rangle$\footnote{
All three triplets, $|T_{0}\rangle$, 
$|\!\uparrow\uparrow\rangle$,
$|\!\downarrow\downarrow\rangle$,
give the same result.}.
Substituting $|\pm\rangle$ defined in (\ref{entangled_state})
for $|\Psi\rangle$ and using the fact that the unpolarized currents
are invariant when all spins are reversed, the expectation  value
$\langle\pm|\delta I_{\alpha} \delta I_{\beta}|\pm\rangle$
can be expressed as the sum of a direct and an exchange term,
\begin{equation}
  \label{cross4}
  \langle\pm|\delta I_{\alpha} \delta I_{\beta}|\pm\rangle
  =  \langle\uparrow\downarrow|\delta I_{\alpha} \delta I_{\beta}|
  \uparrow\downarrow\rangle
  \pm\langle\uparrow\downarrow|\delta I_{\alpha} \delta I_{\beta}|
  \downarrow\uparrow\rangle \enspace ,
\end{equation}
where the upper (lower) sign of the exchange term refers to triplet
(singlet).
Direct calculation of
(\ref{cross4}) gives the following result for the 
zero-frequency ($\omega = 0$) correlations, 
\begin{equation}
  S_{\alpha\beta}
   = \frac{e^2}{h\nu}\left[\sum_{\gamma\delta}\!{}^{'}
    A_{\gamma\delta}^{\alpha}A_{\delta\gamma}^{\beta}
   \mp \delta_{\epsilon_1,\epsilon_2}
    \left(A_{12}^{\alpha}A_{21}^{\beta} +A_{21}^{\alpha}A_{12}^{\beta})
\right)\right] \enspace ,
\label{cross5}
\end{equation}
where $\sum_{\gamma\delta}^{\prime}$ denotes the sum over $\gamma=1,2$ and
all $\delta\neq\gamma$.

We apply formula (\ref{cross5}) now to the set-up shown in Fig. \ref{fig}a 
involving
four leads, described by the single-particle scattering matrix elements,
$s_{31}=s_{42}=r$, and $s_{41}=s_{32}=t$,
where $r$ and $t$ denote the reflection and transmission amplitudes at the
beam splitter, respectively.
We assume that there is no backscattering,
$s_{12}=s_{34}=s_{\alpha\alpha}=0$.
The unitarity of the s-matrix implies $|r|^2+|t|^2=1$, and Re$[r^*t]=0$.
Using  (\ref{A2}) and (\ref{cross5}), we obtain the final
result for the noise correlations for the
incident states $|\pm\rangle$,
\begin{eqnarray}
&&  S_{33}=S_{44}=-S_{34}=eF\left|\langle I\rangle\right| \enspace , \nonumber \\
&&  F =  2eT(1-T)\left(1\mp \delta_{\epsilon_1,\epsilon_2}\right) \enspace ,
\label{fano}
\end{eqnarray}
where $\left|\langle I\rangle\right| = e/h\nu$ is the average current in all leads,
$T=|t|^2$ is the probability for transmission through the
beam splitter, and $F$ is the Fano factor.
Note that the total current $\delta I_3 + \delta I_4$ does note fluctuate, i.e.
$S_{33}+S_{44} +2S_{34}=0$, since we have excluded backscattering. 

 Above results (\ref{fano}) imply that if two electrons
with the same energies, $\epsilon_1=\epsilon_2$, in the singlet
state $|S\rangle = |-\rangle$ are injected into the leads $1$ and $2$,
then the zero frequency noise is {\it enhanced} by a factor of two,
$F=4eT(1-T)$, compared to the shot noise of uncorrelated particles,
$F=2eT(1-T)$. This enhancement of noise is
due to {\it bunching} of electrons in the outgoing leads, caused by the
symmetric orbital wavefunction of the spin singlet $|S\rangle$.
On the other hand, all three triplets $|+\rangle$
exhibit an {\it antibunching} effect, leading to a complete
suppression of the zero-frequency noise, $S_{\alpha\alpha}=0$.
The noise enhancement for the singlet $|S\rangle$ is a
unique signature for entanglement (there exists no unentangled state with
the same symmetry), therefore entanglement can be observed by
measuring the noise power of a mesoscopic conductor
as shown in Fig. \ref{fig}a.\footnote{
These results remain valid in the presence of a Fermi sea.}

\section{Probing Entanglement  of Electrons in a Double-Dot}

The DD system (see Fig. \ref{fig}b) contains $4$ metallic leads which are
in equilibrium with associated reservoirs kept at the chemical
potentials $\mu_i$, $i=1,\ldots ,4$.
The leads are weakly coupled to the dots with 
tunneling amplitudes $\Gamma$ and $\tilde{\Gamma}$, and  the leads
$1,2$  are coupled to {\it both} dots and play the role
of probes where the currents $I_i$ are measured.
The leads $3$ and $4$ are feeding
electrodes to manipulate
the electron filling in the dots.
The quantum dots contain one (excess) electron each, and
are coupled to each other by the tunneling amplitude $t$, which
leads to a level splitting \cite{Loss98,Burkard} $J=E_{\rm t}-E_{\rm s}\sim 4t^2/U$ 
in the DD,  with $U$ being the single-dot
Coulomb repulsion energy, and $E_{{\rm s}/{\rm t}}$ are the singlet/triplet
energies.
We recall that for two electrons in the DD
(and for weak magnetic fields) the ground state is given by a spin
singlet.
For convenience we count the chemical potentials $\mu_i$ from $E_{\rm s}$.
The coupling $\tilde{\Gamma}$ to
the feeding leads can be switched off while probing the DD
with a current.
Here we assume that $\tilde{\Gamma }=0$.

Using a standard tunneling Hamiltonian approach \cite{Mahan},
we write
$H=H_0+V$, where  the first term in $H_0=H_{\rm D}+H_1+H_2$  describes
the DD and $H_{1,2}$ the leads
(assumed to be Fermi liquids).
The tunneling between leads  and dots is described by the perturbation
$V=V_1+V_2$,
where
\begin{equation}\label{perturbation}
V_n=\Gamma\sum_s\left[D^{\dag}_{n,s}c_{n,s}+c^{\dag}_{n,s}D_{n,s}\right],
\quad
D_{n,s}=\E^{\pm \I\varphi/4}d_{1,s}+\E^{\mp \I\varphi/4}d_{2,s} \enspace ,
\end{equation}
and where the operators $c_{n,s}$ and $d_{n,s}$, $n=1,2$, annihilate electrons with
spin $s$ in the $n$th lead and in the $n$th dot,  resp.
The Peierls phase $\varphi$ in the hopping
amplitude accounts for an AB or Berry phase (see below) in the
presence of a magnetic field. The upper
sign belongs to lead 1 and the lower  to
lead 2.
Finally, we assume that spin is conserved in the tunneling process.
For the outgoing currents we have
$I_n=\I e\Gamma\sum_s\left[D^{\dag}_{n,s}c_{n,s}-c^{\dag}_{n,s}D_{n,s}\right]$.
The observable of interest is the average current
through the DD system, $I=\langle I_2\rangle$.

From now on we concentrate on the CB regime where we can
neglect double (or higher)  occupancy in each dot for all  transitions
including virtual ones, i.e. we require $\mu_{1,2}<U$.
Further we assume that  $\mu_{1,2} >J, k_{\rm B}T $ to avoid
resonances  which might change the DD state.
The lead-dot coupling $\Gamma$   is assumed to be weak so that
the  state of the DD is not perturbed; this will allow us
to retain only the first non-vanishing contribution in $\Gamma$
to $I$.
Formally, we require  $J> 2\pi\nu_{\rm t} \Gamma^2$,
where
$\nu_{\rm t}$ is the tunneling density of states of the leads.
In analogy to the single-dot case \cite{averinazarov}  we refer to
above CB regime as  cotunneling regime. 

Continuing with our derivation of $I$,  we note that the average
$\langle\ldots\rangle\equiv {\rm T}{\rm r} {\rho}\left\{\ldots\right\}$
is taken with respect to the  equilibrium
state of the {\it entire} system set up in the distant past
before $V$ is switched
on \cite{Mahan}. Then, in the interaction picture,
the current is given by
\begin{equation}\label{current1}
I=\langle U^{\dag}I_2(t)U\rangle ,\quad U=
{T}\exp\left[-\I \int^t_{-\infty} dt' V(t') \right] \enspace .
\end{equation}
The leading contribution in $\Gamma$ to the cotunneling current
involves the tunneling of
one  electron from the DD to, say, lead
2 and of a second electron from lead 1 to the DD (see Fig. \ref{fig}b). This
contribution is of order  $V_2V_1^2$, and thus $I\propto \Gamma^4$, as
is typical for cotunneling \cite{averinazarov}.
Taking the trace over Fermi leads,
we arrive then at the following compact expression for the
cotunneling current
\begin{eqnarray}
&&I=\frac{1}{2}e\pi\nu^2_{\rm t}\Gamma^4\sum_{i,f,s,s'}\rho_i\,
|\langle i|D_{2,s'}^{\dag}D_{1,s}|f\rangle|^2
\frac{ \Delta_{i,f}\theta(\Delta_{i,f})}{\mu_1\mu_2} \enspace , \nonumber \\
&&\Delta_{i,f}=\mu_1-\mu_2 +E_i-E_f \enspace .
\label{current2}
\end{eqnarray}
This equation shows that
in the cotunneling regime the initial state $|i\rangle$ (with weight
$\rho_i$) of the
DD  is changed into a final state $|f\rangle$
by the traversing electron.
However, due to the weak coupling $\Gamma$, the DD will have returned to
its equilibrium state before the next electron passes through it.

For small bias, $|\mu_1-\mu_2|<J$, only elastic cotunneling
is allowed, i.e.  $E_i=E_f$. However, this regime is not
of
interest here since singlet and triplet contributions turn out to be
identical
and thus indistinguishable.
We thus focus on the  opposite regime, $|\mu_1-\mu_2|>J$, where inelastic
cotunneling\footnote{Note that the AB effect is not suppressed by this inelastic 
cotunneling, since the {\it entire} cotunneling process involving also leads is elastic:
the initial and final states of the {\it entire} system have the same
energy.} 
occurs with singlet and triplet contributions
being
different.
In this regime we can neglect the
dynamics generated by $J$ compared to the one generated by the bias (``slow
spins''), and drop the energies $E_i$ and $E_f$ in (\ref{current2}).
Finally, using
${\rm 1}=\sum_f|f\rangle\langle f|$
we obtain
\begin{eqnarray}
&&I=e\pi\nu_{\rm t}^2\Gamma^4C(\varphi ){{\mu_1-\mu_2}\over {\mu_1\mu_2}} 
\enspace ,\\ \label{current3}
&&C(\varphi )=\sum_{s,s'}\left[\langle
d^{\dag}_{1s^{\prime}}d_{1s}d^{\dag}_{1s}d_{1s^{\prime}}\rangle +\cos
\varphi\langle d^{\dag}_{1s^{\prime}}d_{1s}d^{\dag}_{2s}d_{2s^{\prime}}
\rangle \right] \enspace .
\label{factor1}
\end{eqnarray}
For the purpose of our analysis we assume that
the DD is in its ground state.
Equation (\ref{current3})
shows that the cotunneling
current depends on the properties of the ground state of the DD
through the coherence factor $C(\varphi )$ given in  (\ref{factor1}).
The first term in $C$ is the contribution
from the topologically trivial tunneling path which runs from
lead 1 through, say, dot 1 to lead 2 and back. The second
term (phase-coherent part) in $C$ is the ground state amplitude of  the exchange  
of electron
1 with electron 2 via the leads 1 and 2 such that a closed loop is formed
enclosing an area $A$ (see Fig. \ref{fig}b).
Thus, in the presence of a magnetic
field $B$, an AB phase factor  $\varphi=ABe/h$ is acquired.

Next, we evaluate $C(\varphi)$ explicitly
in the singlet-triplet basis.
Note that only the singlet
$|S\rangle$ and the triplet $|T_0\rangle$  (see (\ref{entangled_state}))
are entangled EPR pairs while
the remaining triplets 
$|T_{+}\rangle=|\!\uparrow\uparrow\rangle$, and
$|T_{-}\rangle =|\!\downarrow\downarrow\rangle$
are not (they factorize).
Assuming that the DD is in one of these states we obtain the
important
result
\begin{equation}\label{factor2}
C(\varphi )= 
\left\{\begin{array}{ll}
2-\cos \varphi \, , \quad & \mbox{for singlet}\\
2+\cos \varphi \, , \quad & \mbox{for all triplets}\enspace .
\end{array}\right.
\end{equation}
Thus, we see that the singlet 
and the triplets 
contribute with {\it opposite sign to the phase-coherent part of the
current}.
One has to distinguish, however,
carefully the entangled from the non-entangled states.
The   phase-coherent part of the entangled states is
a genuine {\it two-particle} effect, while the one of the product states
cannot be distinguished from a phase-coherent {\it single-particle}
effect.
Indeed, this follows from the observation that the
phase-coherent part in $C$ factorizes for the product states
$T_{\pm}$ while it does not so for $S, T_0$. Also, for states such as
$|\!\uparrow\downarrow\rangle$ the coherent part of $C$ vanishes, showing that
two
different (and fixed) spin states cannot lead to a phase-coherent
contribution since we {\it know} which electron goes which part of the loop.
Finally we note that due to the AB phase the role of the singlet and
triplets
can be interchanged which is to say that we can continually transmutate the
statistics of the entangled pairs $S,T_0$ from  fermionic to bosonic
(like in anyons): the symmetric orbital
wave function of the singlet $S$ goes into an antisymmetric one at half a
flux
quantum, and vice versa for the triplet $T_0$.

We would like to stress that
the amplitude of the AB oscillations is a direct measure of the phase
coherence of the entanglement, while the period via the enclosed area
$A=h/eB_0$
gives a direct
measure  of the non-locality of the EPR pairs,
with $B_0$ being the field at which $\varphi=1$. The triplets themselves
can  be further distinguished by applying a directionally inhomogeneous
magnetic
field (around the loop) producing a Berry phase $\Phi^{\rm B}$ \cite{LossBerry},
which is
positive (negative) for the triplet $m=1 (-1)$, while  it vanishes for the
EPR
pairs $S, T_0$. Thus, we will eventually see beating in the AB
oscillations due to the positive (negative) shift of the AB phase $\Phi$ by
the
Berry phase,
$\varphi=\Phi \pm \Phi^{\rm B}$.

\section{Transport of Entangled Electrons}

We consider the general scenario of the transport of entangled electrons 
in a mesoscopic system \cite{DiVincenzoLossMMM}. 
In a first step we inject entangled electrons into the leads and create the state
$|\bf 12\rangle\equiv|\pm\rangle$ (see (\ref{entangled_state})) on the top of the Fermi sea
(as discussed e.g. in Sect. (2), see Fig. \ref{fig}a).
In a second step, we perform a quantum measurement of the state.
As a measure of correlations we
consider transition amplitudes between an initial and a final
state. We begin with the simplest case given by the wave function
overlap of $|\bf 12\rangle$ with $|\bf 34\rangle$,
\begin{equation}
\langle \bf 12|\bf 34\rangle=
\delta_{\bf 13}\delta_{\bf 24}\mp\delta_{\bf 14}\delta_{\bf 23} \enspace ,
\label{overlap}
\end{equation}
 where the upper (lower) sign refers to triplet (singlet).
If the quantum numbers coincide, $\bf 1=3$, and $\bf 2=4$, the overlap assumes its maximum
value $1$, reflecting maximum correlation between the two states. 

Next we generalize this concept 
to leads which contain many interacting electrons besides the two
entangled electrons. 
We  use a similar overlap
as a measure of how much weight remains in the final state
$|{\bf 34},  t\rangle$
when we start from some given initial state
$|\bf 12\rangle$.
The overlap (\ref{overlap}) now becomes a triplet-triplet or singlet-singlet
correlation function
\begin{equation}
G^{{\rm t}/{\rm s}}({\bf 12},{\bf 34};t)= -G({\bf 1},t)\, G({\bf 2},t)\, 
\left[\delta_{\bf 13}\delta_{\bf 24}\mp\delta_{\bf 14}\delta_{\bf 23}\right] 
\enspace  ,
\end{equation}
where we have assumed that
there is no interactions between lead 1 and 2. 
Thus the problem is reduced to the evaluation of (time-ordered)
single-particle Green's functions $G({\bf 1},t)$, $G({\bf 2},t)$
pertaining to lead 1 and 2, resp.  
(these leads are still interacting many-body systems though).  

For the special case $t=0$, and no interactions, we have $G=-\I$,
and  thus
$G^{{\rm t}/{\rm s}}$ reduces to the rhs of  (\ref{overlap}). For the general case,
we evaluate $G$ close to the Fermi surface and get the standard result \cite{Fetter}
\begin{equation}
G(\epsilon, t) \approx -\I z_{\epsilon}
\Theta(\epsilon -\epsilon_F) \E^{-\I \epsilon t- \Gamma_{\epsilon}t}\enspace ,
\label{quasiparticlepole}
\end{equation}
where $\epsilon$ is the quasiparticle energy, 
$\epsilon_{\rm F}$ is the Fermi
energy, and $1/\Gamma_{\epsilon}$ is the
quasiparticle lifetime. In a 2DEG,
$\Gamma_{\epsilon}\propto (\epsilon -\epsilon_{\rm F})^2
\log(\epsilon -\epsilon_{\rm F})$  \cite{Quinn} within the random phase
approximation (RPA).
Thus, the
lifetime becomes infinite when the energy of the added electron
approaches $\epsilon_{\rm F}$. 

 Now, we come to the most important quantity in the present context, the
{\em quasiparticle weight}, $z_{\rm F}=z_{\epsilon_{\rm F}}$,
evaluated at the Fermi surface; it is defined by
\begin{equation}
z_{\rm F}=\left[1-  {\partial \over
{ \partial \omega}}Re \Sigma( \epsilon_{\rm F},\omega=0)\right]^{-1}\enspace ,
\end{equation}
where $\Sigma(\epsilon ,\omega)$ is the irreducible self-energy occurring in the
Dyson equation.  The quasiparticle weight, $0\leq z_{\epsilon}\leq 1$, describes
the weight of the bare electron in the quasiparticle state $\epsilon$,
i.e. when we add an electron with energy $\epsilon \geq \epsilon_{\rm F}$
to the system, some weight (given by $1-z_{\epsilon}$) of the original state
$\epsilon$ will be distributed among all the electrons due to the
Coulomb interaction. 

Restricting ourselves now to energies close to the
Fermi surface we have
\begin{equation}
G^{{\rm t}/{\rm s}}({\bf 12},{\bf 34};t)= z_{\rm F}^2
\left[\delta_{\bf 13}\delta_{\bf 24}\mp\delta_{\bf 14}\delta_{\bf 23}\right] \enspace ,
\end{equation}
{\it for all times} satisfying $0< t\la 1/\Gamma_{\varepsilon}$. Thus we see
that it is the
quasiparticle weight squared, $z_{\rm F}^2$, which is the measure
of our spin correlation function $G^{{\rm t}/{\rm s}}$ we were looking for.
It is thus interesting to evaluate $z_{\rm F}$ explicitly. This is
indeed possible, again within
RPA, and we find after careful calculation \cite{LBS}
\begin{equation}
z_{\rm F}=1-r_{\rm s} ({1\over 2} +{1\over \pi})\enspace ,
\label{qpweight}
\end{equation}
in leading order of the interaction parameter $r_{\rm s}=1/q_{\rm F} a_{\rm B}$, 
where $a_{\rm B}=\epsilon_0\hbar^2/me^2$ is the Bohr radius.  In particular, in a
GaAs 2DEG we have $a_{\rm B}=10.3$ nm, and $r_{\rm s}=0.614$, and thus we
obtain from (\ref{qpweight}) the value $z_{\rm F}=0.665$. We note that a more
accurate numerical evaluation of the exact RPA self-energy yields\footnote{
For 3D metallic leads with, say, $r_{\rm s}=2$ (e.g. $r_{\rm s}^{\rm Cu}=2.67$) 
the loss of correlation is somewhat less strong, since then the quasiparticle
weight becomes $z_{\rm F}=0.77$ \protect\cite{Mahan}.}
$z_{\rm F}=0.691$ \cite{LBS}.
Thus, we see that the spin correlation is reduced by a factor
of about two 
as soon as we inject the two
electrons 
into separate leads consisting of {\it
interacting} Fermi liquids in their ground state.

\section{Acknowledgements} 
This work has been supported by the Swiss National Science Foundation.

\end{document}